# Quantum control in nearly non-degenerate qubits interacting with light

J Wolff[1,*], C Newell[1], A Martin[2], J H McGuire[1,2]

[1]Two State Research Group, Pine View School, Osprey, FL 34209
[2]Department of Physics, Tulane University, New Orleans, LA 70118

Emails: josh.wolff.7@gmail.com, criz1971819@gmail.com, amartintpa@gmail.com, mcguire@tulane.edu

**Abstract.** We consider the time evolution of nearly degenerate two-state systems with an external interaction. Conditions in which full population transfer is possible when the two-states become degenerate are considered. A new variation of the coupled differential equations for the probability amplitudes suggests a possible singularity in the time evolution of the system. The singularity occurs at the point in time where full control in the degenerate limit is achieved. We solve this new variation to derive more information about non-degenerate population control. The new results of numerical calculations for occupational probabilities of non-degenerate systems are presented and interpreted. These results are used to understand how population control breaks down as the photon energy and the energy of the system are changed. Applications are discussed, with an emphasis on interactions of twisted-vortex and plane-wave photons with a variety of targets, including macroscopic gas cells of atoms, of large single crystals, and of molecules.

Keywords: photon-atom interactions, orbital angular momentum, qubit, atomic coherence, coherent control, light-matter interactions, scattering theory

PACS: 03.65.Nk, 32.80.-t, 37.10.Jk, 42.50.Ct

## 1. Introduction

Degenerate qubits are simpler than non-degenerate qubits, where the energy transfer between states is non-zero. Full quantum control of the population of the states by an external interaction is relatively easy to formulate in some degenerate qubits [1,2]. Here we consider population control in nearly degenerate qubits and study the effect of the non-degeneracy, $\Delta E$, on population control of nearly degenerate atomic qubits interacting with light. The qubits we consider allow full quantum control in the degenerate limit under certain conditions that rotate the initial state by $\pi/2$ on the surface of the Block sphere that generally describes qubits.

Our results apply to qubits that are currently being studied in degenerate or nearly degenerate systems [3-8]. This applies to quantum computation and quantum information, where qubits form the basis of these areas of study [7,9-12]. A number of qubits can be modeled by a two-state quantum system with nonzero energy differences [13-18], such as those presented in this paper.

In previous, closely related research, analytical solutions for population control have been found for degenerate qubits [1,2,19]. The effects of dynamic electron correlation in the interactions of atoms and molecules with light have been studied in such a degenerate system [1,20]. The interacting light involves both plane-wave and twisted-vortex photons. These studies are related to optical beams used to control atoms and molecules [21-26] and to other many-body, time-dependent problems [19,27-32]. Degenerate and nearly degenerate two-state systems have also been studied where twisted-vortex photon (and electron) interactions with matter might be used to write information at the sub-atomic level [3-5,19,33].

In the Methods section, standard mathematical methods are used to introduce a new variation for the time-dependent Schrödinger equation by transforming the standard first order differential Schrödinger equation to an equation second in time. This enables application of the method of Frobenius [34,35] that is conventionally used to determine the radius of convergence of a power series solution. When non-degeneracy, $\Delta E$, is introduced, a singularity appears at $t = T/4$ in our variation of the Schrödinger equation. Here $T$ is the period of the oscillating electric field of the perturbing interaction – in our case, the field of the photon.

In the Results section, we present numerical solutions for the differential equation for various values of energy difference between the two states, $\Delta E$, found using a standard Runge-Kutta method. We focus on the effects of non-degeneracy in the perturbative regime of $\Delta E$. Applications are considered in the Discussion.

## 2. Methods

We begin with the two-state approximation [22],

$$\psi(\vec{r}, t) = a_{11}(t)\phi_1(\vec{r})e^{iE_1 t/\hbar} + a_{12}(t)\phi_2(\vec{r})e^{iE_2 t/\hbar}, \qquad (1)$$

where $\phi_n(\vec{r})$ are the orbital functions and $a_{ij}(t)$ are the probability amplitudes for a transition from $\phi_1$ to $\phi_2$. In this qubit system, the two atomic qubits are coupled by an external interaction, $H_{int}$, so that a change in the state of one reflects a change in the state of the other. For example, two electronic spin states in an atom are coupled by a photon. In this case, the system itself represents a two-state system.

The time-dependent Schrödinger equation, $i\hbar \frac{d\Psi}{dt} = H\Psi$ with $H = H_o + H_{int}$, directly leads to [1,2],

$$i\hbar \dot{a}_{11}(t) = E_1 a_{11}(t) + H_{12} \cos\left(\frac{2\pi t}{T}\right) a_{12}(t) = H_{12} \cos\left(\frac{2\pi t}{T}\right) a_{12}(t)$$

$$i\hbar \dot{a}_{12}(t) = E_2 a_{12}(t) + H_{12} \cos\left(\frac{2\pi t}{T}\right) a_{11}(t) = \Delta E a_{12}(t) + H_{12} \cos\left(\frac{2\pi t}{T}\right) a_{11}(t), \qquad (2)$$

where $H_{12}$ is the matrix element of the interaction, $H_{int}$, with the two-state target, and $E_n$ are the electronic eigenstates of $H_o$. Here we consider interactions with a simple time dependence corresponding to $<\phi_1|H_{int}|\phi_2> = H_{12}\cos\left(\frac{2\pi t}{T}\right)$, where $H_{12}$ is independent of time. For mathematical simplicity, we take $E_1$ to be the arbitrary zero energy so that $E_2$ is equal to $\Delta E$, the energy difference between the two states. The energy scale is set by $E$, the energy of the photon, namely $E = hf = \frac{h}{T} = \frac{2\pi\hbar}{T}$ so that $\Delta E$ is expressed in the same units as $E$.

It has been shown [1,2] that the probability amplitudes may be expressed analytically by,

$$a_{11}(t) = \cos\left[\frac{H_{12}}{\hbar}T\sin\left(\frac{2\pi t}{T}\right)\right]$$

$$a_{12}(t) = i\sin\left[\frac{H_{12}}{\hbar}T\sin\left(\frac{2\pi t}{T}\right)\right]. \tag{3}$$

Here the electronic state is initially in the eigenstate, $\phi_1$, and $\phi_2$ is unoccupied. By inspection, one can see that full quantum control is possible when $\frac{H_{12}}{\hbar}T = \frac{\pi}{2}$. Since we are interested in population control in nearly degenerate states, we hereafter generally apply the condition $\frac{H_{12}}{\hbar}T = \frac{\pi}{2}$ in our results.

When $\Delta E \neq 0$, there are no obvious analytical solutions. Rather, solutions may be found numerically. While these may be obtained directly from the conventional time dependent Schrödinger equation, it is useful to rewrite the first order coupled Schrödinger equations for this system as second order coupled differential equations in order to show the convergence properties of power series solutions and to detect possible singularities in the probability amplitudes themselves. By taking the time derivatives of (2) above, it is straightforward to obtain,

$$\ddot{a}_{11}(t) + \left(\frac{2\pi}{T}\tan\left(\frac{2\pi t}{T}\right) + i\frac{\Delta E}{\hbar}\right)\dot{a}_{11}(t) + \frac{H_{12}^2}{\hbar^2}\cos^2\left(\frac{2\pi t}{T}\right)a_{11}(t) = 0$$

$$\ddot{a}_{12}(t) + \left(\frac{2\pi}{T}\tan\left(\frac{2\pi t}{T}\right) + \frac{\Delta E}{\hbar}i\right)\dot{a}_{12}(t) + \left(\frac{H_{12}^2}{\hbar^2}\cos^2\left(\frac{2\pi t}{T}\right) + i\frac{2\pi\Delta E}{T\hbar}\tan\left(\frac{2\pi t}{T}\right)\right)a_{12}(t) = 0. \tag{4}$$

It is now clear that when $\Delta E \neq 0$, (4) potentially have singularities at time $t = T/4$ due to the presence of $\tan\left(\frac{2\pi t}{T}\right)$. Convergent solutions are only guaranteed to exist up until that time according to the well-known method of Frobenius [34,35]. It is only guaranteed that in general any power series based solution will diverge at $t = T/4$. Using a first-order Runge-Kutta method, we do not prove nor disprove the presence of a singularity. Since the singularity is strong (namely an essential singularity), it is not obvious whether or not the amplitudes themselves have a singularity at $t = T/4$. That is, the method of Frobenius [34,35] neither proves nor disproves if there is, or if there is not, a singularity at $t = T/4$. Shakov and McGuire [2] found an analytic (non-singular) solution at $t = T/4$ when $\Delta E = 0$ as indicated by (3), and our numerical results confirm this. While our numerical results for $\Delta E \neq 0$ suggest a possibility for a singularity at $t = T/4$, they do not prove the presence of such a singularity. This leaves open the question of the nature of the solutions of (4) when $\Delta E \neq 0$. In the region of convergence, namely for $t < T/4$, (4) may be calculated numerically using a first-order Runge-Kutta method. Next, we present results based on such numerical solutions. We relate our results to previous research and explain the new findings that come from these numerical calculations, and how they relate to the effect of non-degeneracy.

## 3. Results

Before proceeding to our numerical results, we first emphasize that achieving precise transfer of electron populations from one state to another requires special conditions in quantum systems, as one might anticipate from the uncertainty principle. Our model for a coupled, interacting qubit system (e.g. a qubit interacting with a photon) described by (4) does allow control in the degenerate limit where $\Delta E = 0$. But it does so only when the state of the degenerate qubit is rotated by $\pi/2$, as noted in the introduction. This is manifestly evident in (3) for $\Delta E = 0$, where it is obvious that full transfer is seldom achieved unless $\frac{H_{12}T}{\hbar} = \frac{\pi}{2}$ (or an odd multiple thereof). In this paper we focus on results near or at the limit of full population control, when the condition $\frac{H_{12}T}{\hbar} = \frac{\pi}{2}$ is met.

Our numerical results for the occupational probabilities functions, obtained using (4) with $\Delta E = 0$, are presented in figures 1-3. These numerical calculations use a standard Runge-Kutta method. As shown in figure 1 in the degenerate limit, where $\Delta E = 0$, our numerical results agree with previous results [1,2]. For $\Delta E = 0$, our numerical results match the analytic solutions of (3) and numerically converge well past $t = T/4$, despite the presence of singularities in $\tan(\frac{2\pi t}{T})$.

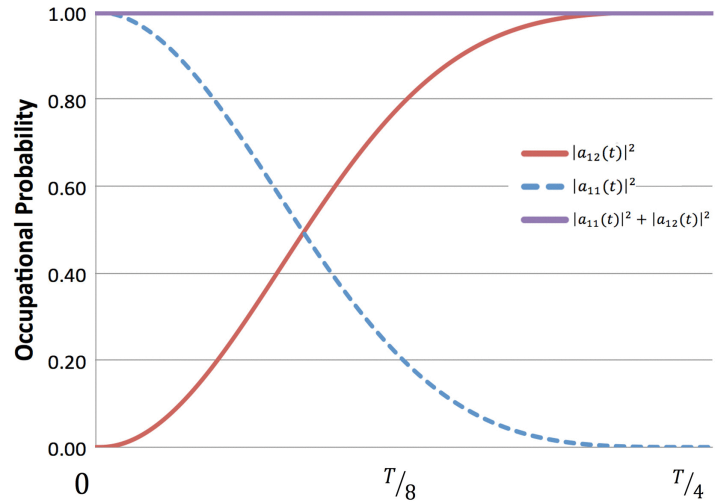

**Figure 1.** Occupational probabilities vs. time. The red and blue lines are the probabilities that the system is in the unoccupied state or the occupied state, respectively. Here $T$ is the period of oscillation of the interaction. The results shown here were calculated numerically from (4) with $\Delta E = 0$. They agree with the analytic results given in (3) for a degenerate qubit system, where $\Delta E = 0$. In both instances, $\frac{H_{12}}{\hbar}T = \frac{\pi}{2}$ is applied. The sum of the two probabilities, shown by the purple line, is equal to 1, consistent with unitarity.

Numerical results for non-degenerate qubits, using $\Delta E \neq 0$ in (4), are shown in figure 2. Nearly full control is observed at $t = T/4$ for small values of $\Delta E$, namely for $10^{-6}E, 10^{-5}E, 10^{-4}E, 10^{-3}E$, and $10^{-2}E$. The numerical solutions for these non-degenerate systems were nearly the same as those for $\Delta E = 0$, differing within 0.012%. Nearly complete quantum control is therefore possible for these nearly degenerate systems. Again, our numerical solutions converged for values of $t$ past $t = T/4$ for small $\Delta E$, but diverged past $t = T/4$ more and more quickly as $\Delta E$ increased.

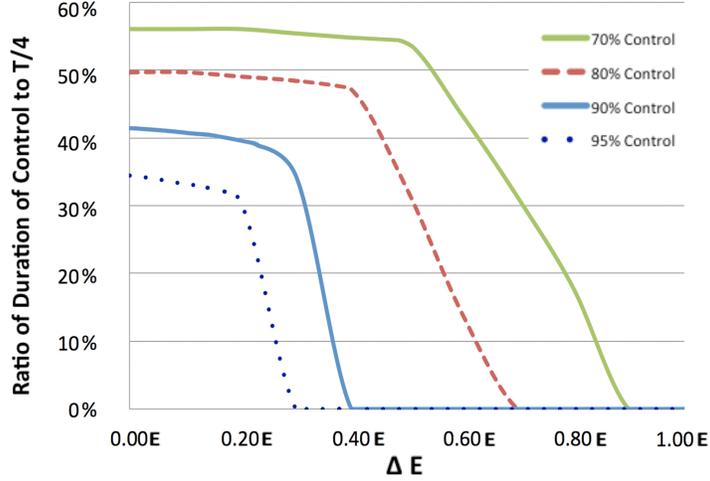

**Figure 2.** Control duration as a function of non-degeneracy, $\Delta E$. The y-axis represents the portion of the first quarter of the period for which there is the indicated population transfer for each $\Delta E$ value. For example, at $\Delta E = 0.20E$, there is 95% population control for approximately 34% of the first quarter of a period. As the $\Delta E$ value increases past the value of $0.10E$, the two-state system enters the non-perturbative regime. Quantum control thereafter depreciates significantly, represented by the sudden, steep slopes in the graph. Note that the condition for quantum control in the degenerate limit, namely $\frac{H_{12}}{\hbar}T = \frac{\pi}{2}$ is applied here.

Quantum control for the time evolution functions begins to break down when $\Delta E > 0.10E$. Figure 2 illustrates this breakdown, where the systems achieve an increasingly small population transfer as the value for $\Delta E$ increases. On the x-axis are values for $\Delta E$. On the y-axis is the percentage of a fourth of a period for which the occupational probability density function $a_{12}(t)$ has the indicated population transfer. The population transfer levels of $95\%, 90\%, 80\%$, and $70\%$ were illustrated. The duration of a period for which the photon is within a certain regime of control begins to gradually lessen and then rapidly decays to 0% as $\Delta E$ increases between $0.10E$ and $1.00E$. At $\Delta E = 0.40E$, quantum control never exceeds 90% for half of a period. At $\Delta E = 0.70E$, quantum control never exceeds 80% in the first fourth of a period. Our data suggests that useful degrees of quantum control for these two-state systems is not limited to degenerate systems, yet is limited to the systems that are nearly degenerate, which are in a perturbative regime. In the non-perturbative regime, where $\Delta E/E \approx 1$, numerical breakdowns of the probability amplitudes occur near $t = T/4$, apparently due to the essential singularity $i\frac{2\pi\Delta E}{T\hbar}\tan(\frac{2\pi t}{T})$ in (4) and/or a possible lack of time symmetry when $\Delta E \neq 0$.

We calculated the effect of non-degeneracy by evaluating the ratio $\left|\frac{|a_{12}(t)_{\Delta E}|^2 - |a_{12}(t)_{\Delta E=0}|^2}{|a_{12}(t)_{\Delta E=0}|^2}\right|$ as a percentage. For $\Delta E = 10^{-2}E$, the effect of non-degeneracy is comparable to the numerical error, yet still existent and parabolic. For $\Delta E = 10^{-1}E$, the effect of non-degeneracy is not small compared to the numerical error, as shown in figure 3.

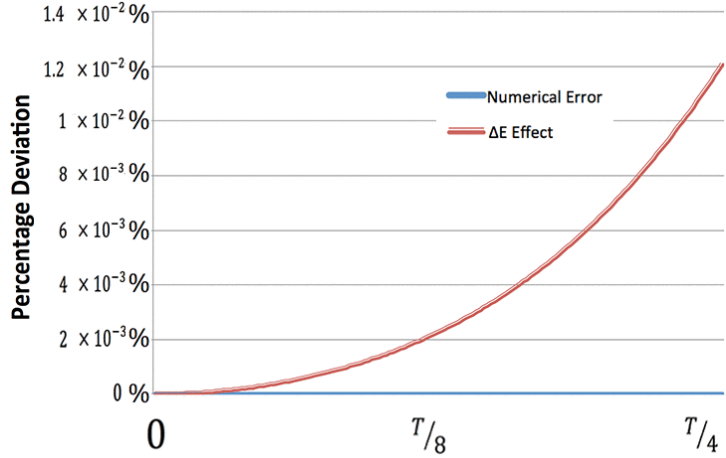

**Figure 3.** Effect of non-degeneracy vs. time for $\Delta E = 10^{-1}E$. The effect of non-degeneracy is the deviation from the degenerate limit presented in figure 1 and (3). The effect of non-degeneracy, shown in red, is essentially parabolic in time. Numerical error, shown in blue, is nonzero yet small compared to the scale of the y-axis.

Analysis of our numerical results provided two points that may be interesting. First, the effect of non-degeneracy is proportional to $(\Delta E/E)^2$ when $\Delta E \leq 10^{-1}E$. This pattern of decay has been reported by other researchers, who also discovered the same proportionality to $(\Delta E/E)^2$ [2]. Our new research adds onto this by finding that the effect of non-degeneracy equals $(\Delta E/E)^2 f(t)$ where f(t) can be estimated by a polynomial function where the odd terms are negative and the even terms are positive. Second, we did a numerical fit to the effect of non-degeneracy for $a_{12}(t)$ when $\Delta E = 10^{-2}E$, yielding $f(t) \approx c_1 t^2 + c_2 t^3 + c_3 t^4$ where $c_1 = 3.24 \times 10^{-1}$, $c_2 = -4.76 \times 10^{-3}$, and $c_3 = 6.89 \times 10^{-2}$. The second term may contain information about higher order perturbation effects.

## 4. Discussion

The quantum description of degenerate and non-degenerate two-state systems where $\frac{H_{12}}{\hbar}T = \frac{\pi}{2}$ holds can be extended to practical applications at the atomic and the molecular level. This includes control of systems in which light interacts with matter [16,36,37], quantum computation and information [12,27,38], and quantum control [13,14,39,40].

In principle, our results may be applied to qubits interacting with both plane-wave and more complex twisted-vortex photons, with wavelengths ranging from radio waves to hard x-rays. Twisted-vortex photons in the visible regime were first identified in 1936 by Beth [41]. New mathematical descriptions of twisted-vortex photons beginning in 1992 [42] stimulated experimental studies involving orbital angular momentum not present in simpler plane-wave photon beams [30]. More recently, mathematical descriptions of twisted-vortex photons interacting with atoms [1,3-5,23] now make it possible to evaluate a cross sections and reaction rates for a broad range of atomic targets. Geometric structure factors enable these calculations to be applied to crystals and molecules [1]. Moreover, relatively simple twist factors can now be used to convert cross sections for plane-wave photons to those for twisted-vortex photons (and electrons) [36]. However, to our knowledge it is not known how commonly reactions involving twisted vortex photons occur in nature (e.g., the earth's atmosphere and the universe) where they could affect our understanding of various natural processes. Some reactions can be quite different for twisted-vortex photons than for plane-wave photons. For example, there is an enhanced effect for elastic (degenerate) Compton scattering in the forward direction for twisted-vortex photons, where a parity restoring double mirror effect removes a parity violating restriction present for plane-wave photons [19].

A simple illustration of the interaction of a twisted-vortex photon is shown in figure 4. In degenerate systems, the occupational probability function for the transfer from one state to the second has a nearly flat distribution in time, indicating nearly full population transfer (See figure 1). A specific example is controlling population transfer of electrons to different electron or molecular bonding orbitals, such as within the 2p-orbital for the case of twisted vortex photons. Here, atomic orbital angular momentum may be exchanged with photon orbital angular momentum [35,43,44]. Figure 4 illustrates an example of forward scattering of the photon beam with the atomic nucleus at the center of the photon vortex on a microscopic scale (with x-ray photons) [1].

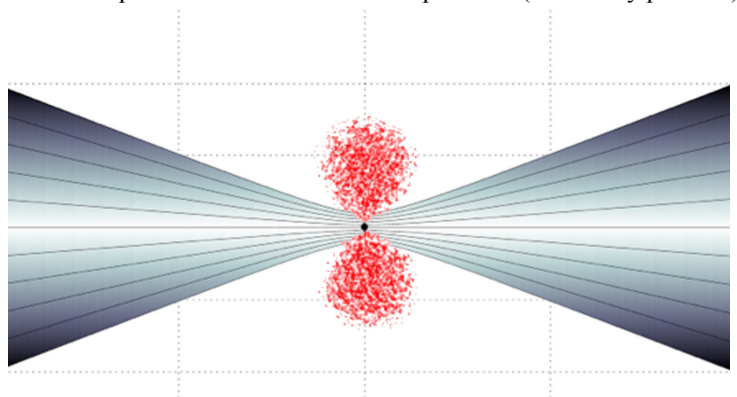

**Figure 4.** Atomic view of a twisted vortex beam of photons interacts with a 2p electronic atomic orbital. The manipulation of the orbital angular momentum of the twisted vortex beam creates a two-state system through which population transfer of within a two-state system can be induced. This creates a qubit capable of storing information [1,19]. In this illustration, the atom is centered at the center of the vortex of the photon and the photon is scattered into the forward direction.

Recent developments offer promise of a wide range of future applications involving the interaction of light (within, and possibly outside, the visible region) with macroscopic gas targets, crystals, large molecules, and, perhaps, even biological units [1,19,36,37,45,46].

## 5. Summary

In this paper, we have detailed the effect of non-degeneracy, $\Delta E$, in population control of a two-state, nearly degenerate system where the specific ratio $\frac{H_{12}}{\hbar} T = \frac{\pi}{2}$ was applied, which is required for full quantum control in the degenerate limit. A variation of the coupled differential probability amplitudes of the Schrödinger equation for this two-state system in the Frobenius limit uncovered an essential singularity in the differential equations for the probability amplitudes at $t = T/4$, the time at which full population control occurs in the limit of degeneracy. Numerical calculations were used to detail limitations of quantum control when non-degeneracy is introduced. The applicability of these results to systems of interaction between light and matter with a broad range of targets is discussed.

**Acknowledgements**

We acknowledge Kh. Shakov for his contribution to the development of the computer code to solve the differential equations. We thank A. Wilson for sponsoring this research.